\documentclass[aps,prb,twocolumn,groupedaddress,showpacs]{revtex4-1}
\usepackage[pdftex]{graphicx}
\usepackage{dcolumn}
\usepackage{bm}
\usepackage{amsmath}
\usepackage{txfonts}
\usepackage{url}
\usepackage{float}
\usepackage{booktabs}
\usepackage[usenames,dvipsnames]{color}
\usepackage{tabularx}
\definecolor{Gray}{gray}{0.0}
\definecolor{lightGray}{gray}{0.35}

\begin{document}
\title{Machine learning clustering technique applied to 
powder X-ray diffraction patterns to distinguish alloy substitutions}
\author{
  Keishu Utimula$^1$,
  Rutchapon Hunkao$^2$, 
Masao Yano$^3$, 
Hiroyuki Kimoto$^3$, 
Kenta Hongo$^{4,5,6,9}$, 
Shogo Kawaguchi$^{7}$, 
Sujin Suwanna$^1$, 
Ryo Maezono$^{8,9}$ \\}

\affiliation{\\
  $^1$School of Materials Science, JAIST, 
  Asahidai 1-1, Nomi, Ishikawa 923-1292, Japan\\
  \\
  $^2$Optical and Quantum Physics Laboratory,
    Department of Physics, faculty of Science,
    Mahidol University, Bangkok 10400, Thailand\\
  \\
  $^3$Advanced Material Engineering Div., 
        Toyota Motor Corporation, 
        Toyota-cho 1, Toyota, Aichi 471-8572, Japan\\
        \\
  $^4$Research Center for Advanced Computing 
        Infrastructure, JAIST, Asahidai 1-1, Nomi, 
        Ishikawa 923-1292, Japan\\
        \\        
  $^5$Center for Materials Research by Information Integration,
  Research and Services Division of Materials Data
  and Integrated System,
  National Institute for Materials Science,
  Tsukuba 305-0047, Japan\\
  \\
  $^6$PRESTO, JST, Kawaguchi, Saitama 332-0012, Japan\\
  \\
  $^7$Japan Synchrotron Radiation Research Institute (JASRI)
  Sayo-gun, Hyogo 679-5148, Japan\\
  \\
  $^8$School of Information Science, Japan Advanced Institute of Science
  and Technology (JAIST),
  Asahidai 1-1, Nomi, Ishikawa 923-1292, Japan\\
  \\
  $^9$Computational Engineering Applications Unit,
    RIKEN, 2-1 Hirosawa,
    Wako, Saitama 351-0198, Japan
}

\begin{abstract}
We applied the clustering technique using 
DTW~(dynamic time wrapping) analysis 
to XRD~(X-ray diffraction) spectrum patterns 
in order to 
identify the microscopic structures 
of substituents introduced in the main 
phase of magnetic alloys. 
The clustering is found to perform well 
to identify the concentrations of the 
substituents with successful rates ($\sim$90\%). 
The sufficient performance is attributed to 
the nature of DTW processing to filter out 
irrelevant informations such as 
the peak intensities (due to the 
incontrollability of diffraction conditions 
in polycrystalline samples) and 
the uniform shift of peak positions 
(due to the thermal expansions of lattices).
The established framework is 
applicable not limited to the system 
treated in this work but widely to 
the systems to be tuned their 
properties by atomic 
substitutions within a phase. 
The framework has larger potential 
to predict wider properties from 
observed XRD patterns in a way that 
it can provide such properties 
evaluated from predicted microscopic local structure, 
such as magnetic moments, optical spectrum  {\it etc.}). 
\end{abstract}
\maketitle

\section{Introduction}
\label{sec.intro}
The concept of {\it materials informatics} 
based on the {\it big data science} 
has attracted recent interests in 
the context for discovering and exploring 
novel materials \cite{Ikebata2017}. 
Achieving high efficiency to get {\it data}, 
namely experimental measurements and analysis 
of materials, is necessary to accelerate the 
cycle of the exploration. 
XRD (X-ray diffraction) analysis is quite commonly 
used to capture crystal structures causing
material properties~\cite{Hongo2018}. 
The analysis is getting accelerated 
by the improvements in X-ray intensities 
as well as in the environments of measurements~\cite{Kawaguchi2017}. 
Typical efforts to achieve efficiency in 
the analysis include such studies applying 
machine learning technique to 
series of XRD data in a systematic observation 
({\it e.g.}, dependences on concentrations, 
temperature {\it etc.}) 
to extract significant information~\cite{Park2017}.
While materials informatics approaches combined with XRD data 
have been recently used to
distinguish different phases (i.e., {\it inter-phase} identifications)
~\cite{2014KUE, 2017SUR, Iwasaki2017, 2018SHA, 2018STA, 2018XIN, 2018OSE},
no attempt has been made to tackle {\it intra-phase} ones so far.
Along this context, 
the present study aims to provide a framework 
which can predict the concentrations of 
atomic substituents introduced in the main phase 
of polycrystalline magnetic alloys. 

\begin{figure}[htb]
\begin{center}
  \includegraphics[scale=0.4]{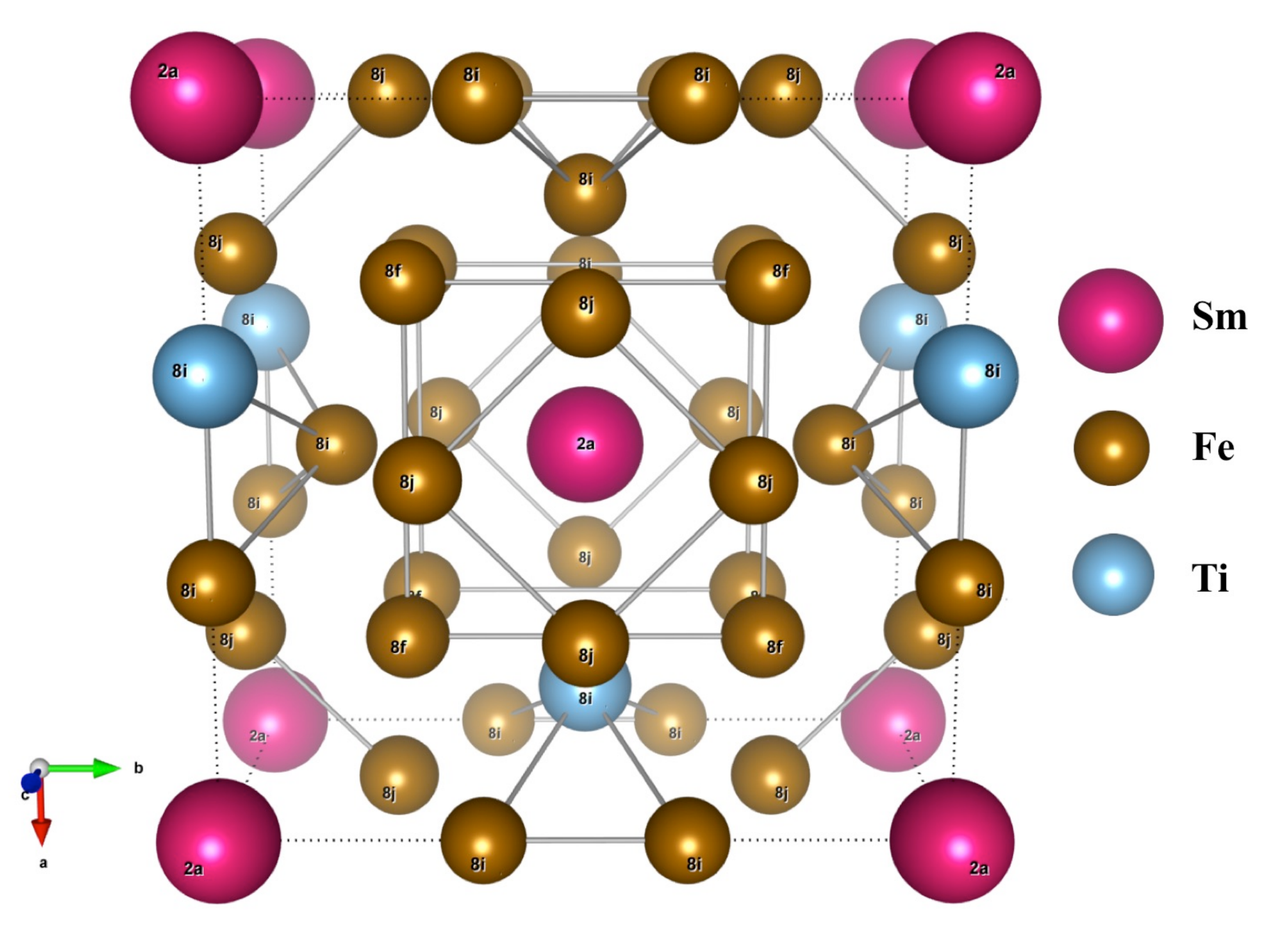}
  \end{center}
\caption{
  The tetragonal ($Imm2$) crystal structure of
  SmFe$_{11}$Ti. Note that the labels are Wyckoff sites of
  space group before substitution by Ti ($I4/mmm$).}
\label{fig_cystalstructure}
\end{figure}

\vspace{2mm}
ThMn$_{12}$-type~(Fig.~\ref{fig_cystalstructure}) crystal
structured SmFe$_{11-x}$Ti$_x$
has been regarded as 
one of the candidates of the main phase in 
rare-earth permanent magnets~\cite{KOBAYASHI2017}.
The origin of intrinsic properties emerging at high temperature 
as well as that of the phase stability has not yet been 
clarified well.
Introducing Ti and Zr to substitute Fe and Sm is found to 
improve the magnetic properties and the phase stability,
as described in details in Sec.'Samples and experiments'. 
To clarify the mechanism how the substitutions improve the properties, 
it is desirable to identify substituted sites and its amount quantitatively, 
preferably with high throughput efficiency 
for accelerating the materials tuning. 
In this work, 
we have developed a machine learning clustering technique 
to distinguish powder XRD patterns to get such microscopic 
identifications about the atomic substitutions. 

\vspace{2mm}
{\it Ab initio} calculations are used to generate 
supervising references for the machine learning of XRD patterns: 
We prepared several possible model structures with substituents 
located on different sites over a range of substitution fractions. 
Geometrical optimizations for each model give slightly different 
structures from each other. 
Then we generated many XRD patterns calculated from each structure. 
We found that the DTW (dynamic time wrapping) analysis 
can capture slight shifts in XRD peak positions corresponding to 
the differences of each relaxed structure, distinguishing 
the fractions and positions of substituents. 
We have established such a clustering technique using 
Ward's analysis on top of the DTW, being
capable of sorting
out simulated XRD patterns based on the distinction. 

\vspace{2mm}
The established technique can hence learn the correspondence 
between XRD peak shifts and microscopic structures with 
substitutions over many supervising simulated data. 
Since the {\it ab initio} simulation can also give several 
properties such as magnetization for each structure, 
the correspondence in the machine learning can further predict 
functional properties of materials when it is applied 
to the experimental XRD patterns, not only being
capable of distinguishing
the atomic substitutions. 
The machine learning technique for XRD patterns developed here 
has therefore the wider range of applications not limited 
only on magnets, but further on those materials which properties 
are tuned by the atomic substitutions.

\section{Results}
\label{sec.results}
For our target system, 
[Sm$_{(1-y)}$Zr$_y$]~Fe$_{12-x}$Ti$_x$, 
we examined the range for $x$ and $y$ 
as shown in Table~\ref{table.searching}, 
which is accessible by the experiments. 
For a given concentration, several possible 
configurations for substituents exist. 
They are sorted into identical subgroups 
in terms of the crystalline symmetries, 
as described in Sec.'Computational details'. 
Table.~\ref{table.SmZr} and ~\ref{table.FeTi} 
summarize the possible space groups 
of substituted alloy structure 
(used as initial structures for computations) 
for given concentrations of 
Sm/Zr and Fe/Ti, respectively. 
\begin{table}[h]
  \caption{
    The numbers of inequivalent configurations of
    Sm$_{(1-y)}$Zr$_y$Fe$_{12-x}$Ti$_x$ to be considered.  
    The numbers in bracket indicate the structure constructed
    from $2\times2\times2$ supercell (Sm/Zr),
    while the rest from  $2\times2\times1$ supercell (Fe/Ti).}
\begin{center}
  \begin{tabular}{c|rrrrr}
    $y\backslash x$ & 0.0 & 0.5 & 1.0 & 1.5 & 2.0 \\
    \hline
    0.000 & 1 & 13 & 22 (1) & 27 & 61 \\
    0.125 & -  & -  & (2) & - & - \\
    0.250 & -  & -   & (7) &- & - \\
    0.375 & -  & -   & (6) &- & - \\
    0.500 & -  & -   & (10) &- &-  \\
  \end{tabular}
\end{center}
\label{table.searching}
\end{table}
\begin{table}[t]
  \caption{[table.SmZr]
    Space groups of the initial structures 
for the substitution models (Sm/Zr) at 
each concentration. For $y=0.5$, for instance, 
inequivalent configurations of 
substituted sites amounts to six in total.
The number given in parenthesis represents the 
number of degenerated configurations within 
each symmetry at the initial structures for 
further lattice relaxations. 
It amounts therefore 26 configurations in total 
for generating simulated XRD patterns. 
  }
  \begin{center}
    \begin{tabular}{llllll}
      \toprule
      &\multicolumn{5}{c}{ $y$ ( space group/number of configurations)}\\
      \hline
      & 0.000 & 0.125 & 0.250  & 0.375 & 0.500 \\
      \midrule
      &\textit{Imm2} (1) & \textit{Imm2} (2) & \textit{Amm2} (4)
      & \textit{Imm2} (2) & \textit{Imm2} (1) \\
      & & & \textit{Cmm2} (2) & \textit{Cm} (2) & \textit{Ima2} (2) \\
      & & & \textit{P1} (1) & \textit{C2} (2) & \textit{Cm} (2) \\
      & & & & & \textit{C2} (2) \\
      & & & & & \textit{Pmm2} (2) \\
      & & & & & \textit{P1} (1) \\
     \bottomrule
    \end{tabular}
  \end{center}
  \label{table.SmZr}
\end{table}
\begin{table}[t]
\caption{Space group of SmFe$_{12-x}$Ti$_{x}$ with
    inequivalent site of Ti substitutions. The SmFe$_{12}$($I4/mmm$)
    is used as an initial structure.
The number given in parenthesis represents the 
number of degenerated configurations within 
each symmetry at the initial structures for 
further lattice relaxations. 
It amounts therefore 124 configurations in total 
for generating simulated XRD patterns. 
}
\begin{center}
\begin{tabular}{llllll}
  \toprule
  &\multicolumn{5}{c}{ $x$ ( space group/number of configurations)}\\
  \hline
  & 0.0 & 0.5 & 1.0 & 1.5 & 2.0 \\
  \midrule
  & \textit{I4/mmm} (1) &\textit{Cm} (1) & \textit{Imm2} (2)
  & \textit{Cm} (2) & \textit{Fmmm} (2) \\
  &            & \textit{C2} (4) & \textit{C2/m} (12) & \textit{C2} (8)
  & \textit{Immm} (4) \\
  &            & \textit{P-1} (8) & \textit{C2/c} (6) & \textit{P-1} (16)
  & \textit{Fmm2} (1) \\
  &            &        & \textit{Cm} (1)    & \textit{P1} (1)
  & \textit{Imm2} (1) \\
  &            &        & \textit{P1} (1)    &        & \textit{C2/m} (20) \\
  &            &        &           &        & \textit{C2/c} (4) \\
  &            &        &           &        & \textit{Cm} (4) \\
  &            &        &           &        & \textit{Cc} (1) \\
  &            &        &           &        & \textit{C2} (8) \\
  &            &        &           &        & \textit{P-1} (14) \\
  &            &        &           &        & \textit{P1} (2) \\
  \bottomrule
\end{tabular}
\end{center}
\label{table.FeTi}
\end{table}

\vspace{2mm}
After applying lattice relaxations 
to the initial structures achieved 
by {\it ab initio} geometrical 
optimizations, we can calculate 
XRD patterns for the lattices. 
The procedures in details for the above 
are given in Sec.'Computational details'. 
We could therefore generate
'simulated XRD patterns' as above, 
{\it e.g.}, 26 patterns for the Sm/Zr substitution, 
those are the data for the clustering by the 
unsupervised learning. 
We examine whether the clustering can 
sort them again correctly based on 
their concentration. 

\begin{figure}[htb]
\begin{center}
  \includegraphics[width=\linewidth]{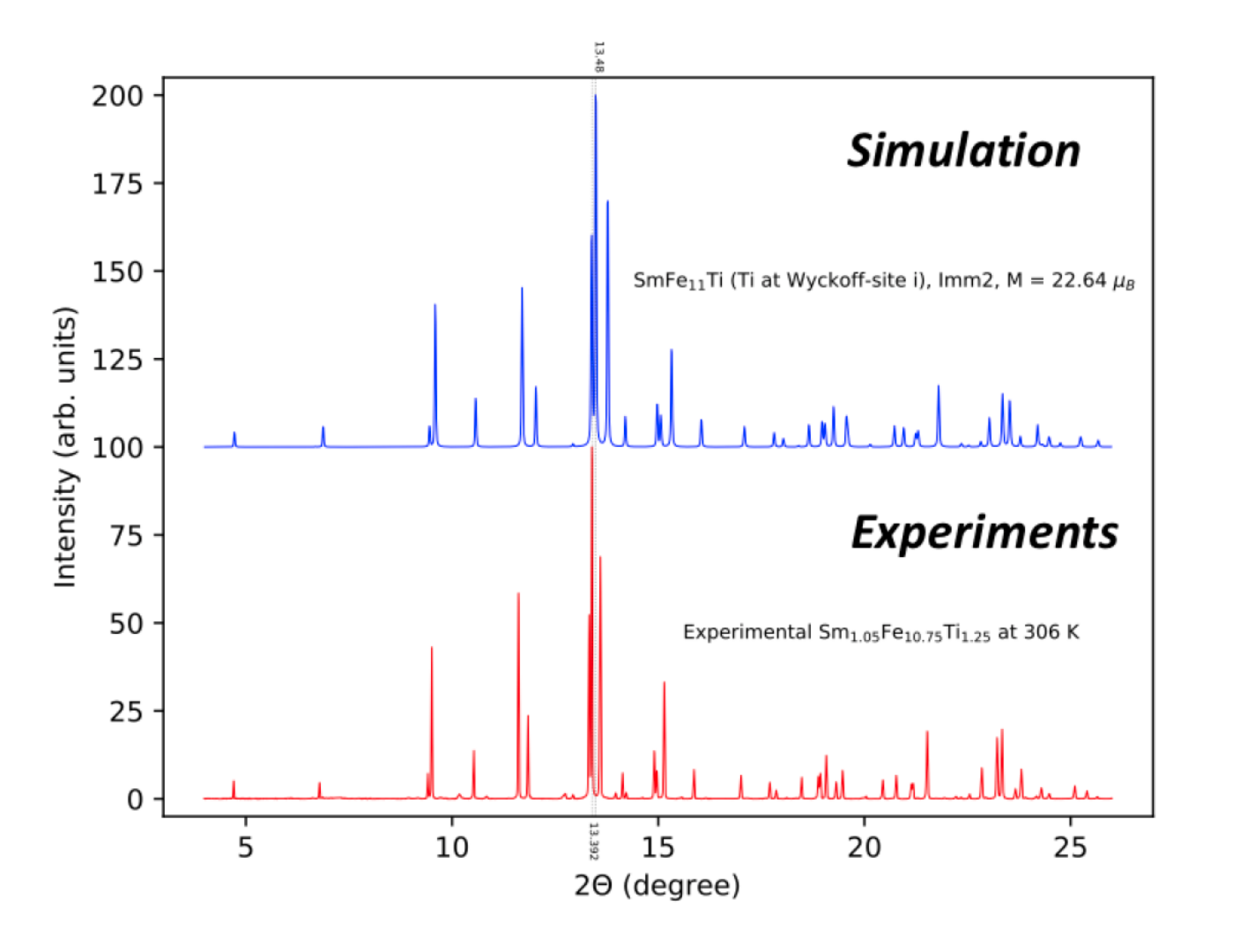}
  \end{center}
\caption{
  Comparison of simulated XRD patterns (bottom) of
  SmFe$_{11}$Ti
  and experimental XRD patterns (top) of
  Sm$_{1.05}$Fe$_{10.75}$Ti$_{1.25}$.
  The inset numbers are the main-phase peak position.}
\label{fig.xrdComparison}
\end{figure}

\vspace{2mm}
Resultant XRD patterns (simulated one)
fairly well coincides 
with experimental ones, as shown in 
Fig.~\ref{fig.xrdComparison}. 
We see that the patterns keep the overall shape 
almost completely, just with slight variations 
in the inter-peak distances depending on 
the concentrations.
To capture only such slight variations, 
DTW is expected to perform well due to the following reason: 
The method is designed to be applied to such signals given 
along an axis ({\it e.g.}, time dependent signal, $y(t)$) 
so that it can extract only the {\it shape} of the signal 
ignoring uniform shifts along the axis.
The method scores the {\it dissimilarity} 
between signals, $i$ and $j$, in terms of the 
DTW-distance, DTW$(i,j)$. 

\begin{figure}[htb]
\begin{center}
  \includegraphics[scale=0.4]{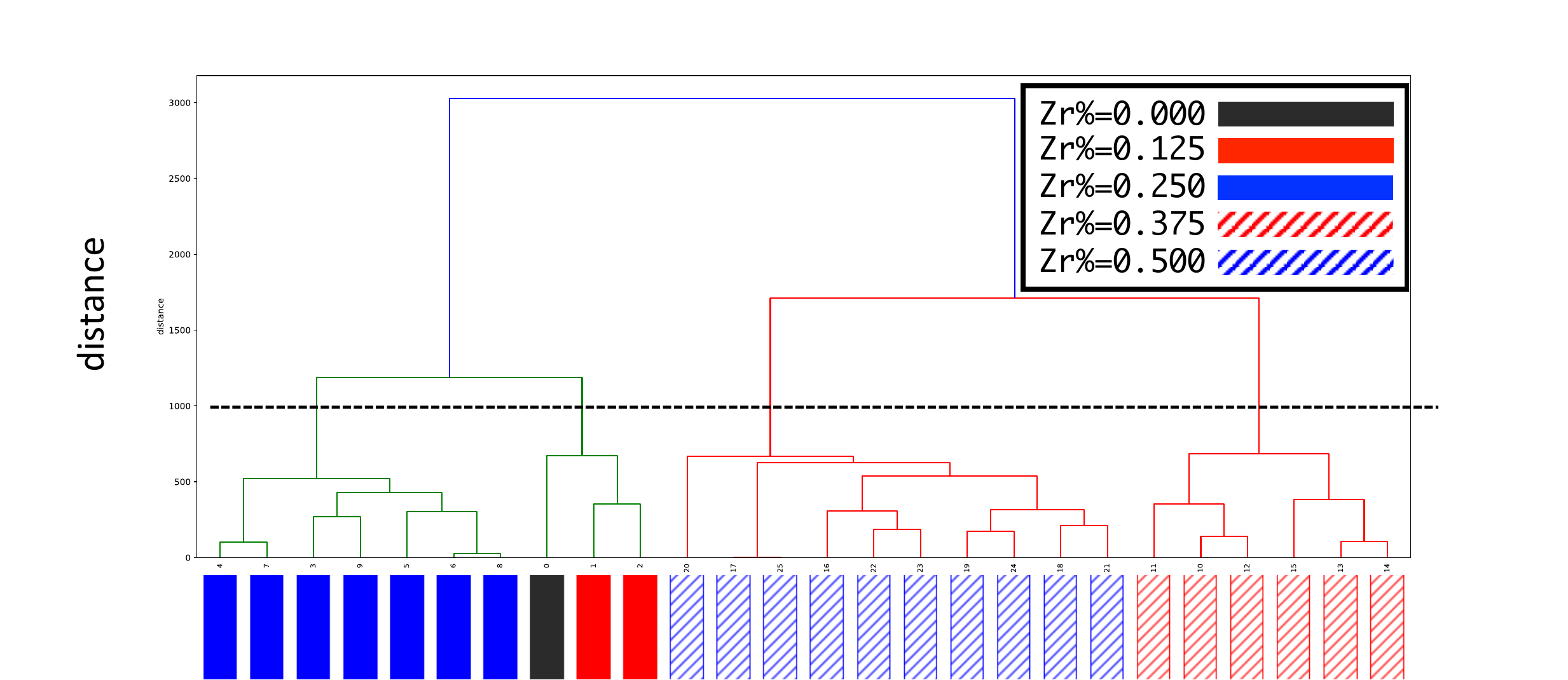}
  \end{center}
\caption{
Clustering over the XRD peak patterns~(26 in total) of 
[Sm$_{1-y}$~Zr$_y$]~Fe$_{11}$Ti, performed by DTW 
(dynamical-time-wrapping) scoring and Ward linkage method. 
Putting the threshold around 1,000 for the dissimilarity 
(horizontal broken line), the patterns are clustered into 
four groups, sharing almost  the same number of the 
substitutions by Zr. 
The red arrows at the bottom show the errors where 
a 'zero substitution' is wrongly sorted into the group 
with 'one substitution' {\it etc.}
}
\label{fig.SmZrClustering}
\end{figure}

\vspace{2mm}
Specifying a clustering framework is generally 
given by a combination of methods, $a\otimes b$, 
where '$a$/that scoring the dissimilarity', 
and '$b$/that making
linkages between 
elements to form clusters based on 
the given dissimilarity'. 
In the present work, we employed the 
framework, [Normalized Constrained DTW (NC-DTW)]
~$\otimes$~[Ward linkage method], 
using those implemented in 'Scipy package'~\cite{Scipy}. 
The descriptions of linkage and dissimilarity-measure 
methods being used in this work can be found on the 
Scipy documents, except the DTW dissimilarity measures
which were calculated by fastDTW\cite{Salvador2007} package. 
The framework is found to achieve the clustering 
to distinguish the concentration of Sm/Zr substitutions 
with sufficiently high accuracy, 96.2\% 
(one failures among 26 XRD patterns), 
as shown in Fig.~\ref{fig.SmZrClustering}. 

\section{Discussions}
\subsection{Limitation of the DTW-dissimilarity}
\begin{figure}[htb]
\begin{center}
  \includegraphics[scale=0.4]{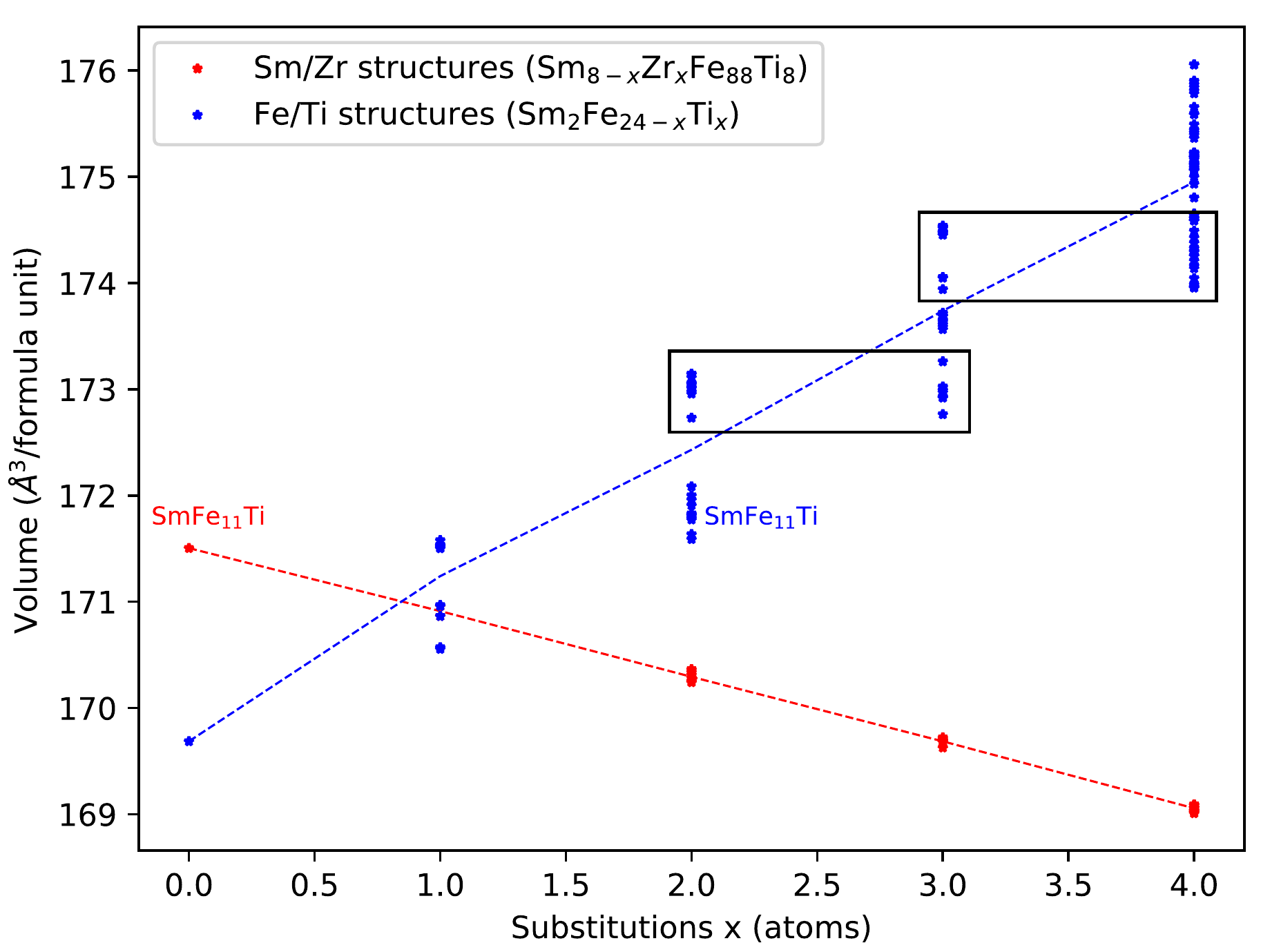}
  \end{center}
\caption{
Dependences of the unit cell volume ($v$) 
on the concentration $x$ for 
Fe/Ti~(blue) and Sm/Zr~(red) substitutions. 
Several plots with the same color on 
the same $x$ has the different symmetries 
as given in Table~\ref{table.SmZr} and \ref{table.FeTi}. 
Rectangular enclosures on the blue dependence 
show the {\it degeneracy}, {\it i.e.}, 
the different $x$ may give the same $v$. 
}
\label{fig.degenerate}
\end{figure}

When the same method (Ward$\otimes$DTW) 
as in Sm/Zr case is applied to Fe/Ti, 
the successful rate for the recognition 
gets reduced to 33.1\%. 
We can identify the reason why the successful 
rate for the Fe/Ti gets worse than that for Sm/Zr 
from the dependence shown in Fig.~\ref{fig.degenerate}. 
Since XRD reflects lattice constants as its peak 
position, we can take the unit cell volume, $v$, 
as a representative quantity to be captured 
by the clustering recognition under such a situation 
where the cell symmetry is kept unchanged. 
DTW dissimilarity, DTW$(i,j)$, can then be 
regarded to be scaling roughly to the 
difference of $v$. 
The recognition can therefore 
be regarded as such 
a framework to perform an {\it inverse inference} 
from the 'difference of $v$' to identify 
the 'difference of $x$' on the dependence of $v(x)$, 
as shown in Fig.~\ref{fig.degenerate}.
For Sm/Zr, the 'trace back mapping' from $v$ 
to $x$ is one-to-one, while for Fe/Ti it is not the case 
due to the {\it degeneracy} in the sense that 
many different values of $v$ share the same $x$. 
Under such a {\it degeneracy}, it is impossible 
to provide correct inferences of 'difference in $x$' 
from a given 'difference in $v$'. 
Such a difficulty occurring for the Fe/Ti case 
leads to the worse successful rate for the 
clustering recognition. 

\begin{figure}[htb]
  \centering
  \includegraphics[width=0.45\textwidth]{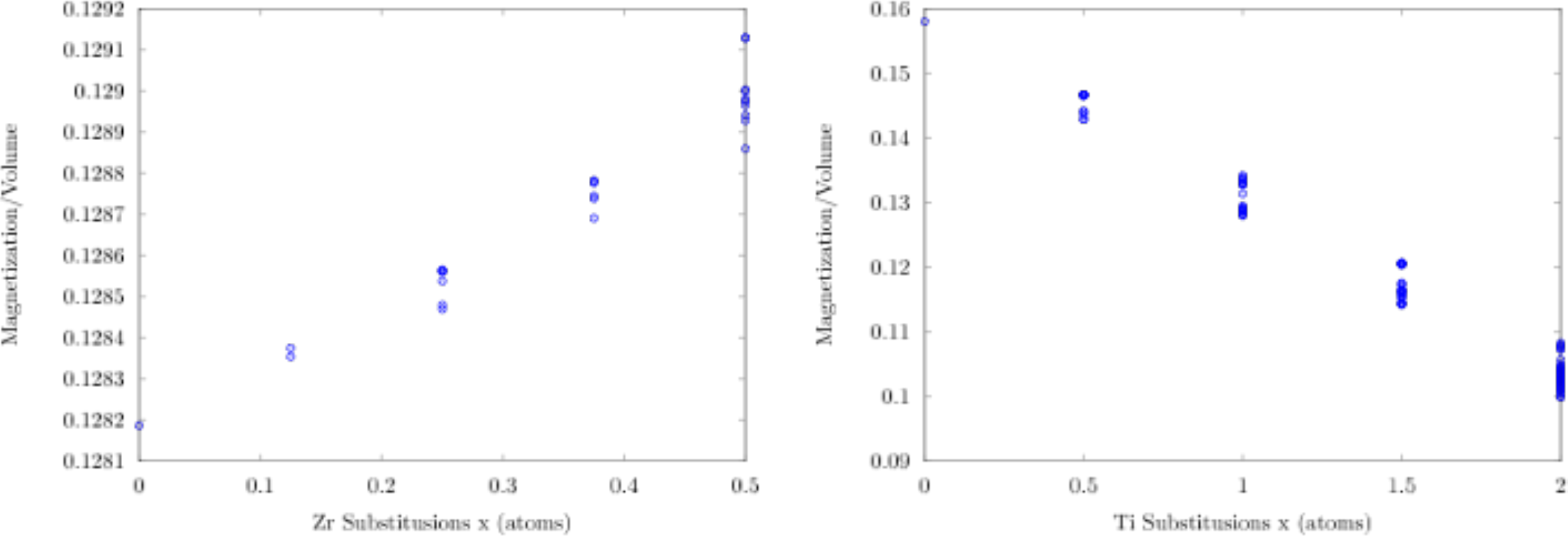}
  \caption{
    Magnetizations depending on
  the concentrations for Sm/Zr (Sm$_{8-x}$Zr$_x$Fe$_{88}$Ti$_8$) [left panel]
  and Fe/Ti structures (Sm$_2$Fe$_{24-x}$Ti$_x$) [right panel]. 
}
\label{fig.mag}
\end{figure}

\begin{figure}[htb]
\begin{center}
  \includegraphics[scale=0.32]{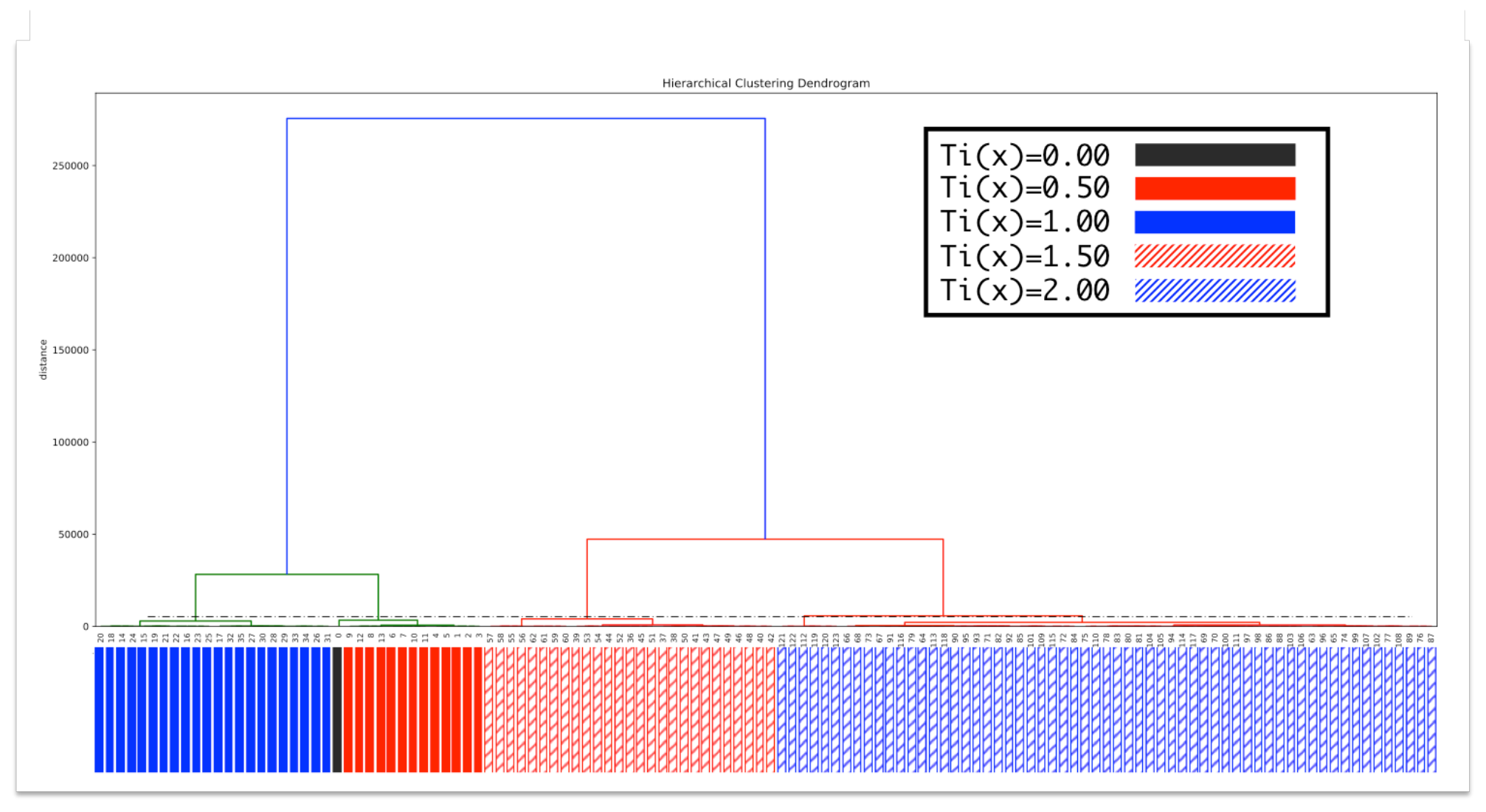}
  \end{center}
\caption{
Clustering over the XRD peak patterns~(124 in total) of 
SmFe$_{12-x}$Ti$_{x}$, performed by DTW 
(dynamical-time-wrapping) scoring and Ward linkage method. 
The weighted function calculated from magnetization was used
to improve the dissimilarity measures.
}
\label{fig.FeTiWeight}
\end{figure}

\vspace{2mm}
The problem can be resolved by a way 
assisted by the advantage of {\it ab initio} 
methods in the sense that they can 
provide several other quantities not only 
the optimized lattice parameters. 
Even when ${\rm DTW}(i,j)\sim \left|{v(x_i)-v(x_j)}\right|$
does not work well due to the degeneracy in $v(x)$, 
other quantities such as the magnetization $M(v)$ 
can be non-degenerate (as shown in Fig.~\ref{fig.mag}) 
and hence useful to solve the difficulty. 
Using magnetizations is especially practical 
because the quantity is available from both 
experiments and simulations. 
We also note that the dependence in 
Fig.~\ref{fig.degenerate} (left panel) is 
consistent with the experimental fact~\cite{2016KUN}
that the magnetization per volume is increasing 
as the Zr concentration increases. 
\begin{equation}
{\rm Dissimilarity}(i,j) = {\rm DTW}(i,j)\times W(i,j) \ ,
\label{improvedW}
\end{equation}
so that it can prevent from the problem 
due to the degeneracy.
We have confirmed that the successful 
rate is actually improved from 33.1\% into 99.19\%,
as shown in Fig. \ref{fig.FeTiWeight},
by using the weight as above. 

\subsection{How to treat experimental XRD}
As shown in Fig.~\ref{fig.xrdComparison}, 
simulated XRD patterns ($s$) well reproduce 
the experimental ones ($e$). 
The consistency is sufficiently 
enough so that the {\it direct} 
comparison to evaluate the DTW 
distance, DTW$(e,s)$, can 
make sense for the clustering 
(it is usual that some pre-processing 
for law data, '$e$' or '$s$', to get 
corrections, '$\tilde e$' or '$\tilde s$'
to evaluate DTW$(\tilde e,\tilde s)$ 
in order to fill the gap between 
the idealized simulations and realities). 
By preparing simulated XRDs, 
$(\left\{ s_j\right\}_{j=1}^{N})$, in advance, 
we can identify such a $s_k$ for a given $e$ 
which gives the smallest distance, 
$\left|e-s_k \right|$. 
The simulated $s_k$ is accompanied by 
several quantities, $\left\{q_\alpha \right\}$, 
such as the formation energy, 
the magnetization, and the local geometrical 
configuration of substituents, those evaluated 
by {\it ab initio} method. 
Then the $\left\{q_\alpha \right\}$ can be 
the theoretical predictions for the observed $e$, 
serving a machine-leaning framework for XRD 
patterns assisted by {\it ab initio} simulations. 
\begin{figure}[htb]
\begin{center}
  \includegraphics[scale=0.4]{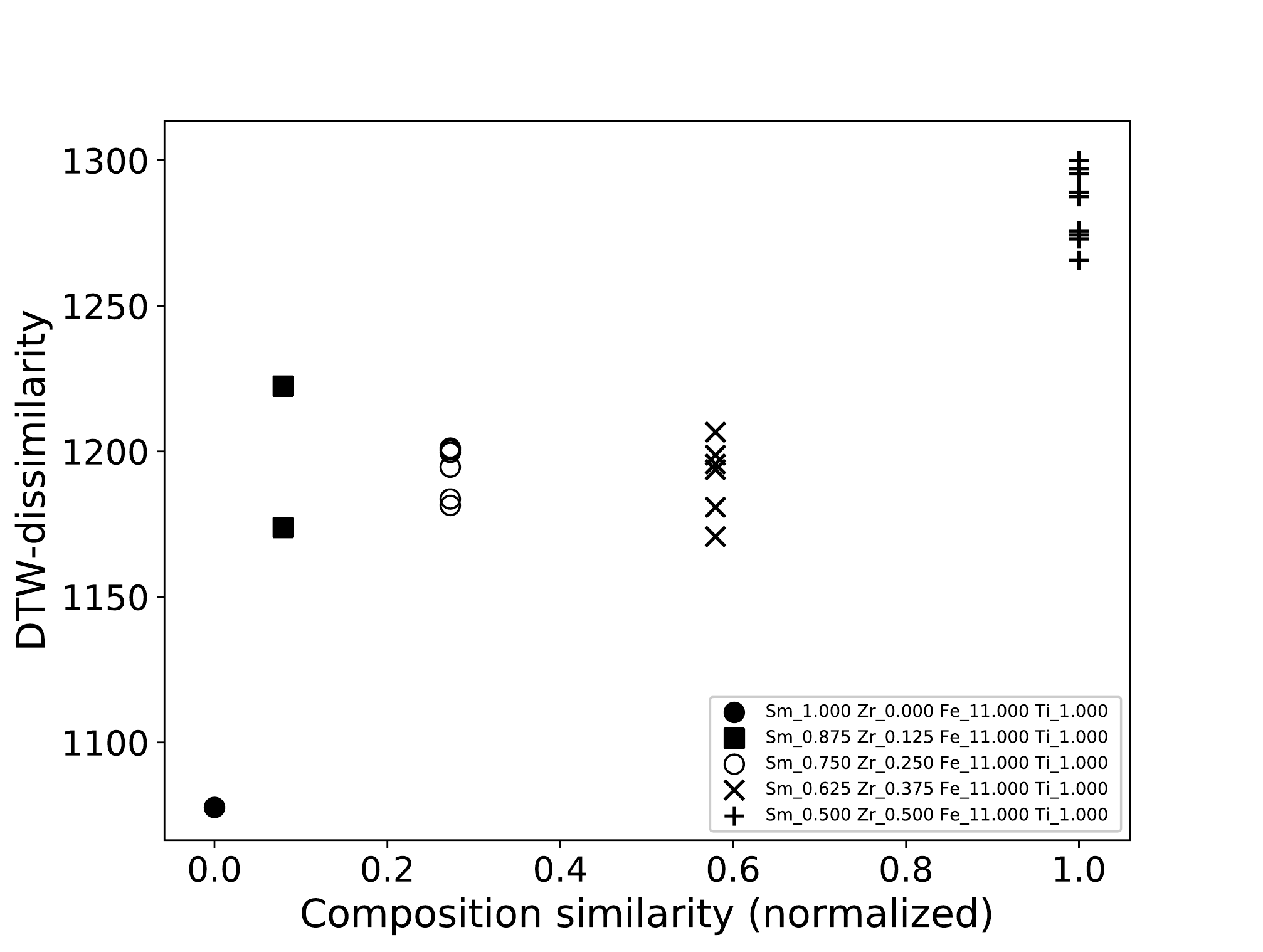}
  \end{center}
\caption{
DTW-dissimilarities 
between 
an experimental XRD ($e$) 
and simulated XRDs ($\left\{ s_j\right\}$), 
being in a correlation with 
the composition similarity.
The $e$ is taken at the composition, 
Sm$_{1.05}$Zr$_{0.0}$Fe$_{10.75}$Ti$_{1.25}$, 
compared with $\left\{ s_j\right\}$ 
in terms of DTW$(e,s_j)$ (vertical variable). 
The horizontal variable is 
the normalized 'composition similarity' 
defined by Eq.~(\ref{CompDist}). 
}
\label{esDist}
\end{figure}
Fig.~\ref{esDist} shows that 
such a distance, $\left|e-s_k \right|$, 
works fairly well, 
taking an example of $e$ 
at a composition 
Sm$_{1.05}$Zr$_{0.0}$Fe$_{10.75}$Ti$_{1.25}$. 
For the general composition, 
Sm$_{c_1}$Zr$_{c_2}$Fe$_{c_3}$Ti$_{c_4}$, 
we can define the 'composition similarity' 
between $e$ and $\left\{ s_j\right\}$ as, 
\begin{equation}
D = \sum\limits_{\alpha  = 1}^4 
{{{\left( {c_\alpha ^{\left( e \right)} 
- c_\alpha ^{\left( s \right)}} \right)}^2}} \ . 
\label{CompDist}
\end{equation}
In Fig.~\ref{esDist}, we see 
that DTW$(e,s_j)$ (vertical variable) 
well correlates with the 'composition similarity'. 
The closest $s_k$ giving the shortest 
DTW$(e,s_k)$ (black filled circle in the figure) 
has actually the closest composition, 
Sm$_{1.0}$Zr$_{0.0}$Fe$_{11.0}$Ti$_{1.0}$ 
than the other $\left\{ s_j \right\}$. 
The prediction accuracy is simply improved 
by the more number of simulation data, 
$(\left\{ s_j\right\}_{j=1}^{N})$. 
A straightforward way to do so is to take more 
dense grid on $x$ but it requires 
larger supercell and hence more computational 
power.
For the present grid resolution, 
the experimental XRD with 
Zr\% = 0.0, 10.4, and 31.8 are 
identified to be closest to 
Zr\% (simulated) = 0, 25, and 37.5, 
respectively, being the best performance 
as possible.

\begin{figure}[htb]
\begin{center}
  \includegraphics[scale=0.4]{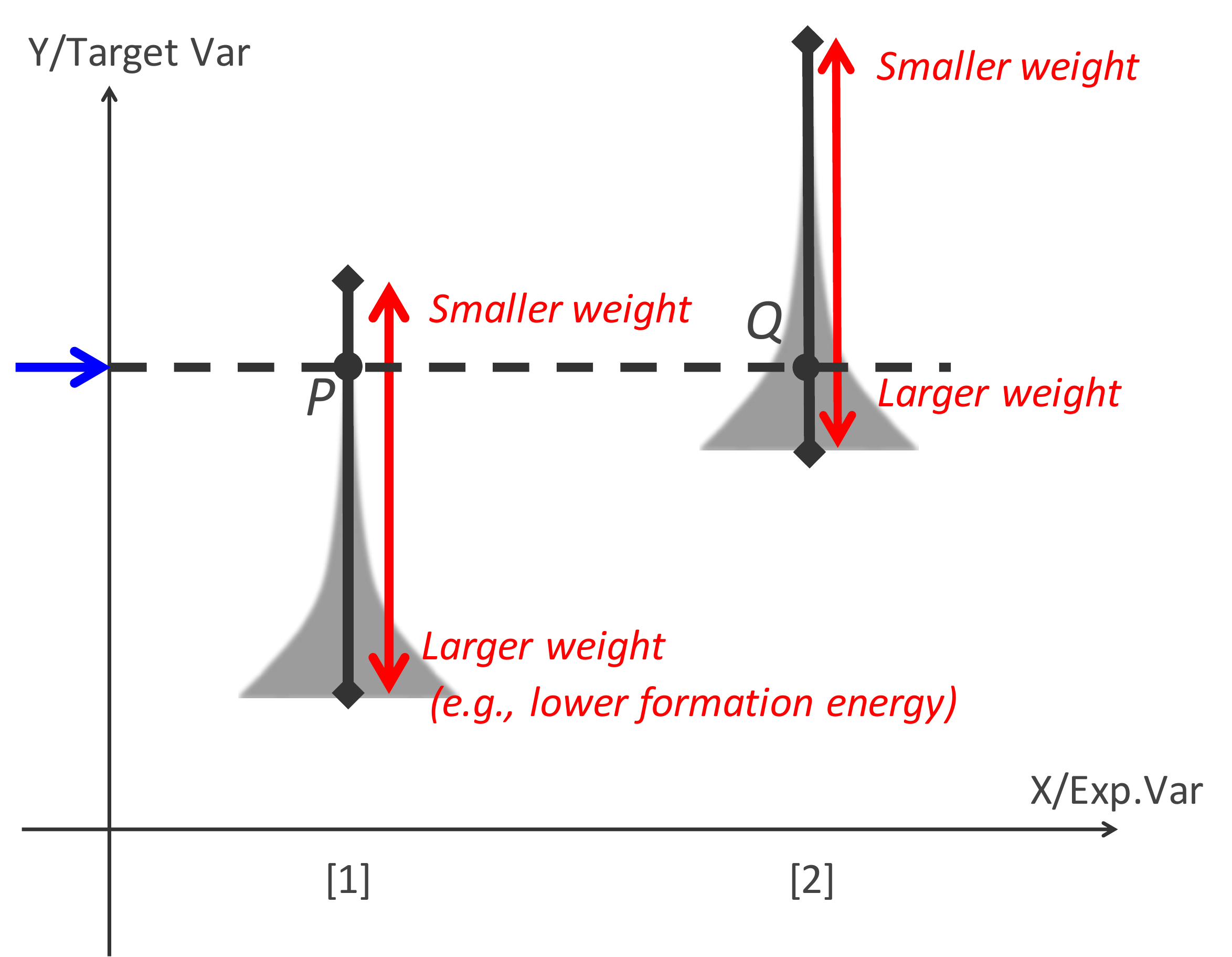}
  \end{center}
\caption{
  A schematic picture explaining the 
difficulty to make proper distinctions 
in the clustering due to the 'degeneracy'. 
The horizontal axis corresponds to 
the substitution concentrations 
in our case, while the vertical one 
to the lattice parameters 
those characterize a XRD pattern. 
The 'errorbar-like' symbols represent 
the data spreading along the vertical axis 
shown in Fig.~\ref{fig.degenerate}. 
When a XRD pattern is given as a 
point on the vertical axis (blue arrow), 
several points ($P$ and $Q$) 
corespond to it with almost the same 
lattice parameters but with different 
internal alignments of defects, 
leading to the difficulty 
to mix up the possibilities of [1] and [2] 
as the possible 'explanation variable 
(concentration in this case)'. 
Red arrows beside the 'errorbars' mean that 
there is the weight, such as the formation energy, 
being possible to put over the spreading. 
}
\label{tempF8}
\end{figure}

\vspace{2mm}
In the case with the degeneracy (Fig.~\ref{fig.degenerate} 
for Fe/Ti substitution), 
the DTW distance is not
capable of performing 
the clustering 
for $\left\{ s_j \right\}$, and hence quite unlikely 
to be 
capable of identifying
the closest $s_k$ for a given 
$e$ based on the $\left|e-s_k \right|$.
The strategy with $W(i,j)$ (the weight by the 
magnetization) introduced in the previous 
section
won't work in this case 
because for $e$ (experimental XRD patterns), 
the accompanying 
quantities such as maginetizations 
are not always available. 
A possible remedy to distinguish $e$ 
would be as follows using a plucked set, 
$\tilde A \subset A=\left\{ s_j \right\}$: 
Since $A$ is generated by simulations, 
each element is accompanied with the 
quantities like the magnetization, 
the formation energy {\it etc.}. 
By using the formation energies, 
we can pluck the degenerating candidates 
({\it e.g.}, $P$ and $Q$ in Fig.~\ref{tempF8}) 
by excluding ones with higher 
energies ($P$ in Fig.~\ref{tempF8}) 
to form the plucked subset $\tilde A$. 
The degeneracy is now excluded 
in $\tilde A$, and hence used 
as a pool of references to be 
identified as the closest $s_k$ to a given $e$ 
based on the DTW distance, $\left|e-s_k \right|$. 
The identified $s_k$ is accompanied with 
the physical quantities evaluated by the simulations, 
and hence they could be the estimates for the 
sample giving the experimental XRD, $e$. 

\subsection{Significance to use DTW}
Clustering package 'Scipy'~\cite{Scipy}
used here includes several other 
algorithms than our DTW$\otimes$Ward choice. 
It is worth interesting to see the comparisons 
of their performance, as shown in 
Table.~\ref{table.score1}-\ref{table.score2} 
with detailed explanations given in \S\ref{SI}.D. 
Although the NC-DTW does not show the best performance there, 
however, the compromise to the peak-shift of NC-DTW is required
when we consider to treat experimental data. 
We note that the simulated XRDs are reflecting 
structures at zero temperature while 
the experimental ones are subject to 
thermal effects under the finite temperature. 
The effects would lead the broadening of peaks 
due to the thermal vibrations as well as 
the peak shifts due to the thermal expansions. 
Since we are looking at the change within a phase 
(not inter-phase changes), the shifts are expected 
to be almost uniform, not so modifying the inter-peak 
distances significantly because the expansion occurs 
almost evenly for every lattice degrees of freedom.
Such uniform shifts are not detected by DTW 
as its intention of the design, and hence 
the scoring works well not affected by the 
thermal effect. 
This nature forms the robustness against 
thermal noises on the experimental data 
enabling the direct comparison with 
simulation data at zero temperature 
to evaluate $\left|e-s_k\right|$. 
Based on the above observations, 
we positively use DTW even though 
it does not achieve the best performance 
for simulated data as seen in 
Table.~\ref{table.score1}-\ref{table.score2}. 
The evidence for this issue has also been shown in 
the preceding study~\cite{Iwasaki2017} as NC-DTW shows
the best performance, among 
various techniques, to sort out the various phases
from experimental data.

\vspace{2mm}
Several preceding works are found 
those applied DTW to analyse XRD patterns
~\cite{Baumes2008,Iwasaki2017}. 
While these studies applied it 
to distinguish phases ({\it i.e.}, 
inter-phase works), the present study 
works on the {\it intra-phase} 
identifications. 
In the formers, DTW is used to 
distinguish {\it major} differences 
of peak positions those drastically 
occurring when the phase changes~\cite{Iwasaki2017}. 
In this study, on the other hand, 
we clarified the new capability of DTW, 
namely, it can distinguish even far tiny 
changes of inter-peak distances 
those occurring within a target phase. 
By this capability, we can explore 
a new framework that enables to identify 
the microscopic geometries of the substituents 
introduced in a target phase assisted 
by machine learning technique. 

\section{Conclusion}
\label{sec.conc}
We have developed such a clustering framework 
that can be applied to XRD patterns of 
alloys to distinguish the concentrations 
of substituents. 
We found that the clustering works 
quite well to identify the concentrations 
when applied to the patterns of 
magnetic alloys based on SmFe$_{12}$. 
Supercell models for the substitutions 
are found to work well with 
{\it ab initio} lattice relaxations, 
reproducing XRD patterns being sufficiently 
in coincidence with experiments. 
The implementation of the clustering 
with [DTW dissimilarity scoring]$\otimes$
[Ward linkage method] is found to 
achieve around 90\% of the successful rate 
for distinguishing the concentration. 
The main reason of the failure case in 
the clustering is identified being due to 
the {\it degeneracy}, namely the situation 
where different concentrations give 
almost the same lattice constant. 
By imposing quantities predicted by 
{\it ab initio} methods into the weight 
used for the dissimilarity scoring, 
such degeneracies are lifted to prevent 
the clustering from failure. 
Sufficiently good coincidence between 
simulated and experimental XRD patterns 
enables the framework to be used 
to predict unknown concentrations 
of the substituent 
introduced in the main phase of alloys 
from their XRD patterns.   
The established framework here is 
applicable not only to the system 
treated in this work but widely to 
the systems to be tuned their 
properties by atomic 
substitutions within a phase. 
Not only identifying the concentrations, 
the framework has larger potential 
for usefulness to predict wider properties from 
observed XRD patterns in a way that 
it can provide such properties 
evaluated from predicted microscopic local structure 
(positions of substitutions {\it etc.}), 
including magnetic moments, optical spectrum  {\it etc.}). 

\section{Acknowledgments}
The computations in this work have been performed 
using the facilities of Research Center for Advanced Computing 
Infrastructure at JAIST.
R.M. is grateful for financial supports from 
MEXT-KAKENHI (17H05478 and 16KK0097), 
from FLAGSHIP2020 (project nos. hp180206 and hp180175 at K-computer), 
from Toyota Motor Corporation, from I-O DATA Foundation, 
and from the Air Force Office of Scientific Research 
(AFOSR-AOARD/FA2386-17-1-4049). 
K.H. is grateful for financial supports from 
FLAGSHIP2020 (project nos. hp180206 and hp180175 at K-computer),
KAKENHI grant (17K17762), 
a Grant-in-Aid for Scientific Research on Innovative Areas (16H06439),
PRESTO (JPMJPR16NA) and
the ``Materials research by Information Integration Initiative" (MI$^2$I) 
project of the Support Program for Starting Up Innovation Hub 
from Japan Science and Technology Agency (JST).
R.H. is grateful for financial support from the Development and Promotion of Science and Technology Talents Project (DPST) for a scholarship to study at Faculty of Science, Mahidol University, and research internship at JAIST.

\section{Supplemental Information}
\label{SI}

\subsection{Samples and Experiments}
The X-ray diffraction (XRD) measurements for the powdered Sm-Fe-Ti were 
performed at the beamline BL02B2 in SPring-8
(Proposal Nos. 2016B1618 and 2017A1602).
CeO$_2$ diffraction pattern was used to determine 
the X-ray energy of 25~keV.
The diffraction intensities were collected using 
a sample rotator system and a high-resolution 
one-dimensional semiconductor detector (multiple MYTHEN system) 
with a step size of 2$\theta$ = 0.006~[deg.]~\cite{Kawaguchi2017}. 
The samples were powderized from strip-casted 
alloys and the powder was put into a quartz capillary 
and encapsulated with negative pressure of Ar gas.

\subsection{Computational Details}
For getting the structures of the target alloys, 
[Sm$_{(1-y)}$Zr$_y$]~Fe$_{12-x}$Ti$_x$, 
we firstly constructed a tetragonal ($I4/mmm$)
crystal structure of SmFe$_{12}$ using 
experimental lattice parameters, $a=0.856$ nm and
$b=0.480$ nm ($b=a$), of SmFe$_{11}$Ti~\cite{experiment-lattice}
as an initial setting for further optimizations. 
For Zr-substitutions replacing Sm sites 
(ranging from 1-4 atoms), 
we constructed $2\times2\times2$ supercell, 
containing 104 atoms in the primitive of 
tetragonal ($Imm2$) of SmFe$_{11}$Ti
(Fig.~\ref{fig_cystalstructure}). 
All possible configurations were considered to cover
a randomness of experimental substitutions,
and we ignored some configurations by considering their symmetry
using FINDSYM software~\cite{FINDSYM}.
Finally, we considered only 26 supercells
(Table.~\ref{table.SmZr}) that possess 
different space groups, Wyckoff site occupations.

\vspace{2mm}
For {\it ab initio} calculations, we used the 
spin-polarized density functional theory (DFT) 
implemented in the 'Vienna {\it ab initio} simulation
package~(VASP)'~\cite{VASP1,VASP2,VASP3}. 
For such systems like our target those 
including transition metal and rare earth elements, 
it is generally known that the predictions are 
critically influenced by the choice of 
exchange-correlation (XC) potentials used
in DFT~\cite{Hongo2017,Ichibha2017,Hongo2015,Hongo2013}
.
For the present case, it has been found 
that DFT+$U$ is essentially inevitable 
if we treat $f$-orbitals as the valence
range\cite{Larson2003,YEHIA2008,LIU2011,Pang2009,Cheng2012} 
It has also been found that 
GGA (generalized gradient approximation) works well 
if the 4$f$ is treated as the core range 
described by pseudo potentials
~\cite{GGA-largecore,GGA-largecore2,GGA-largecore3,Ismail2011,Puchala2013}. 
We therefore used 
the revised Perdew-Burke-Erzenhof (RPBE)~\cite{RPBE} 
for the GGA-XC upon the confirmation 
that RPBE improves optimized lattice parameters 
getting closer to experiment ones 
than when PBE~\cite{PBE} used. 
The pseudopotentials based on projected augmented
wave (PAW)~\cite{VASP-PAW} method were used.
The $s$ and $p$ semi-core states are included in valence states,
except Sm, resulting in 12, 16 and 12 valence states
for Zr, Fe and Ti, respectively. 
The structural relaxations were done until a force on each ion
was smaller than 0.01 eV \AA $^{-1}$.
A plane-wave cutoff energy of 400~eV and $5\times5\times5$
Monkhorst-Pack grids were used which was large enough
to give convergence energy.
The lattice relaxations with the above choice 
applied to SmFe$_{11}$Ti is confirmed 
to get the lowest total energy 
with Ti at 8i site, which is consistent 
with experiments~\cite{isite-experiment1,KOBAYASHI2017}
and {\it ab initio} calculations~\cite{isite-simulation1} 
of RFe$_{11}$Ti-type magnetic compounds.
The optimized lattice parameters, $a$ and $c$, 
were 0.851 and 0.473~[nm] which are good agreement 
with the experiments~\cite{experiment-lattice}
These comparisons confirm that our model 
sufficiently reasonable.
With Ti substitution at 8i site, 
$I4/mmm$ space group breaks
and becomes $Imm2$ as shown in Fig. \ref{fig_cystalstructure}.

\subsection{Validation of simulated XRD patterns}
To validate our simulated XRD pattern, 
the simulated XRD patterns were compared to the 
experimental XRD patterns. 
The X-ray diffraction (XRD) patterns of the optimized
structures were
theoretically calculated by the powder diffraction pattern
utility in VESTA\cite{Momma2011} software.
The X-ray wavelength 
of 0.496~\AA was used
as being used in experiment.
The isotropic atomic displacement parameter ($B$)
was set to 1.00 \AA.
The normalized XRD patterns having 2$\theta$ from
1 to 120 degree with 0.01 degree interval was obtained. 

\vspace{2mm}
The simulated XRD pattern of SmFe$_{11}$Ti agrees 
very well with the experimental XRD pattern of 
Sm$_{1.05}$Fe$_{10.75}$Ti$_{1.25}$ 
(Fig.~\ref{fig.xrdComparison}), 
but the main-phase peak position is quite different as 
it is 13.41 deg. in experiment 
while 13.48 deg. in simulation.
This is due to the fact that the peak shift occurs  
if the lattice expands or contracts, and we found 
that the optimized lattice parameters from DFT are 
underestimated which accounts for the difference. 
This underestimated lattice parameters introduce
only systematic shift of peak position 
but their XRD profiles remain unchanged. 
When the Zr concentration increases, 
the main-phase peak position of the simulated XRD patterns 
shifts to larger 2$\theta$, being in accordance with 
experimental results. 

\subsection{Hierarchical clustering analysis}
The hierarchical clustering analysis (HCA) was used to identify
the simulated XRD patterns. 
All clustering analysis were carried out using Scipy
package~\cite{Scipy}.
The descriptions of linkage and dissimilarity-measure methods
being used
in this work can be found on the Scipy documents, except the
DTW dissimilarity measures
which were calculated by fastDTW\cite{Salvador2007} package.
\begin{table*}[htb]
  \caption{[table.score1]Adjusted rand index of clustering result of
    Sm$_{1-y}$Zr$_{y}$Fe$_{11}$Ti (Sm/Zr) structures.}
\begin{center}
  \begin{tabular}{l|cccccccc}
      & Single & Complete & Average & Weighted & Centroid & Median & Ward \\
    \hline
    NC-DTW & 1.00 & 0.80 & 1.00 & 0.82 & 0.82 & 0.82 & 0.91 \\
    Cityblock & 1.00 & 1.00 & 1.00 & 1.00 & 1.00 & 1.00 & 1.00 \\
    Euclidean & 1.00 & 1.00 & 1.00 & 1.00 & 1.00 & 1.00 & 1.00 \\
    Cosine & 1.00 & 1.00 & 1.00 & 1.00 & 1.00 & 1.00 & 1.00 \\
    Correlation & 1.00 & 1.00 & 1.00 & 1.00 & 1.00 & 1.00 & 1.00 \\
  \end{tabular}
\end{center}
\label{table.score1}
\end{table*}
\begin{table*}[htb]
  \caption{[table.score2]Adjusted rand index of clustering result of
    SmFe$_{12-x}$Ti$_{x}$ (Fe/Ti) structures.}
\begin{center}
  \begin{tabular}{l|cccccccc}
      & Single & Complete & Average & Weighted & Centroid & Median & Ward \\
    \hline
    NC-DTW & -0.04 & -0.10 & -0.08 & -0.08 & -0.04 & -0.04 & 0.01 \\
    Cityblock & -0.06 & 0.49 & 0.45 & 0.39 & 0.27 & 0.40 & 0.41 \\
    Euclidean & -0.03 & 0.28 & 0.47 & 0.49 & 0.27 & 0.28 & 0.28 \\
    Cosine & -0.01 & 0.35 & 0.51 & 0.55 & 0.33 & 0.28 & 0.35 \\
    Correlation & -0.01 & 0.34 & 0.51 & 0.55 & 0.33 & 0.33 & 0.35 \\
  \end{tabular}
\end{center}
\label{table.score2}
\end{table*}
The package provides variety of other 
methods than the present choice, 
DTW$\otimes$Ward, as shown in 
Table~\ref{table.score1}-\ref{table.score2}. 
The tables compares the performances 
achieved by the variety of choices 
for the identifications of Sm/Zr and Fe/Ti, 
respectively. 
The performance is evaluated in terms of 
ARI~(adjusted Rand index), which measures the 
similarity between the true labels and
predicted labels with the maximum score and minimum score of
1 and -1, respectively. The ARI calculations have been done by using
'Scikit-learn' package\cite{Pedregosa2011}.
In the tables, several dissimilarity measures, NC-DTW, 
Euclidean, Cityblock, Cosine and Correlation, 
with various linkage methods, Single, Complete, Average, Weighted,
Centroid, Median and Ward, have been compared. 
The perfect score of 1 is reached by all methods,
except NC-DTW, in Sm/Zr structures.
While the best method in Fe/Ti structures are cosine
and correlation with ward linkage with 0.55 score.
The NC-DTW method provides lower performance than
other methods in both structures since
the NC-DTW omits the peak-shift information while the rest are
peak-position based dissimilarity measure.
With NC-DTW dissimilarity measure, the ward linkage method shows
the a good performance, ARI of 0.91 and 0.01 for Sm/Zr and Fe/Ti
structures, respectively, among all linkage methods. 
Therefore, in this work, we will focus on NC-DTW with ward
linkage method.

\bibliography{references}

\end{document}